\documentclass[runningheads]{llncs}

\usepackage{tikz}
\usetikzlibrary{arrows,calc,fit}
\usepackage{scalefnt}
\usepackage{comment}
\usepackage{graphicx}
\usepackage{subfig}
\usepackage{tabularx}
\usepackage{amsmath}
\usepackage{algorithm}
\usepackage{array}
\usepackage[noend]{algpseudocode}
\usepackage{amsfonts}
\usepackage[acronym]{glossaries}
\usepackage[capitalize]{cleveref}
\usepackage[T1]{fontenc}
\def\doi#1{\href{https://doi.org/\detokenize{#1}}{\url{https://doi.org/\detokenize{#1}}}}
\usepackage{listings}
\loadglsentries{glossary.sty}

\makeatletter
\def\BState{\State\hskip-\ALG@thistlm}
\makeatother
\DeclareMathOperator*{\argmin}{argmin}

\usepackage{makecell, multirow}

\newcommand\asteriskfill{\leavevmode\xleaders\hbox{$\ast\ $}\hfill\kern0pt}

\begin{document}
\title{Weakly-supervised Biomechanically-constrained CT/MRI Registration of the Spine}
\titlerunning{Biomechanical CT/MRI spine registration}
\author{Bailiang Jian\inst{1,2} \and
Mohammad Farid Azampour\inst{1}\orcidID{0000-0003-4077-1021} \and
Francesca De Benetti\inst{1,2}\orcidID{0000-0002-5255-4553} \and
Johannes Oberreuter\inst{2,3} \and
Christina Bukas\inst{2,4}\orcidID{0000-0001-9913-8525} \and
Alexandra S. Gersing\inst{5,6}\orcidID{0000-0003-1687-5541} \and
Sarah C. Foreman\inst{5,6}\orcidID{0000-0001-9140-0162} \and
Anna-Sophia Dietrich\inst{5,6} \and
Jon Rischewski\inst{5,6} \and
Jan S. Kirschke\inst{5,6}\orcidID{0000-0002-7557-0003} \and
Nassir Navab\inst{1,7}\orcidID{0000-0002-6032-5611} \and
Thomas Wendler\inst{1,2}\orcidID{0000-0002-0915-0510}
}
\institute{
Chair for Computer Aided Medical Procedures and Augmented Reality, Technische Universit\"{a}t M\"{u}nchen, Garching, Germany, \email{\{bailiang.jian,wendler\}@tum.de}
\and
ScintHealth GmbH, Munich, Germany,
\and
Reply SpA, Munich, Germany,
\and
Helmholtz AI, Helmholtz Zentrum München, Munich, Germany,
\and
Department of Radiology, Technische Universit\"{a}t M\"{u}nchen, Munich, Germany,
\and
Department of Neuroradiology, Technische Universit\"{a}t M\"{u}nchen, Munich, Germany,
\and
Computer Aided Medical Procedures Lab, Laboratory for Computational Sensing+Robotics, Johns Hopkins University, Baltimore, MD, USA
}

%\author{Paper ID: 507}

\authorrunning{Jian et al.}

%\institute{
%\asteriskfill \\ \asteriskfill \\ \asteriskfill \\ \asteriskfill \\ \asteriskfill \\ \asteriskfill
%}

\maketitle              % typeset the header of the contribution
\begin{abstract}
\gls{CT} and \gls{MRI} are two of the most informative modalities in spinal diagnostics and treatment planning. \gls{CT} is useful when analysing bony structures, while \gls{MRI} gives information about the soft tissue. Thus, fusing the information of both modalities can be very beneficial.
Registration is the first step for this fusion. While the soft tissues around the vertebra are deformable, each vertebral body is constrained to move rigidly. We propose a weakly-supervised deep learning framework that preserves the rigidity and the volume of each vertebra while maximizing the accuracy of the registration. To achieve this goal, we introduce anatomy-aware losses for training the network. We specifically design these losses to depend only on the \gls{CT} label maps since automatic vertebra segmentation in \gls{CT} gives more accurate results contrary to \gls{MRI}. We evaluate our method on an in-house dataset of 167 patients. Our results show that adding the anatomy-aware losses increases the plausibility of the inferred transformation while keeping the accuracy untouched. 

\keywords{CT/MRI registration  \and Deep learning image registration \and Biomechanical constraints \and Spine}
\end{abstract}
\section{Introduction}
\gls{CT} and \gls{MRI} images are the most used imaging modalities for understanding the pathologies of the spine and defining the therapeutic approach. Experts need both modalities for proper diagnosis in complex clinical situations as each modality provides different information~\cite{kim2020diagnostic}. \gls{CT} has higher contrast in bony structures and is highly sensitive for fracture detection and other pathologies~\cite{parizel2010trauma,tins2010technical}. 
% For post-operative evaluation, CT is accurate of identifying the location and integrity for the implants, and assessing the success of decompression and detection related compilations~\cite{ghodasara2019postoperative}.
%\gls{MRI} is usually adopted to distinguish ligaments, tendons, and musculature with its high sensitivity to soft tissue. 
On the other hand, \gls{MRI} can be used to detect lesions and tumors of the spinal cord and the intervertebral discs, as well as to evaluate the inner anatomy of the vertebral bodies~\cite{shah2011imaging}.
% Specifically, post-contrast T1-weighted images are compelling for evaluating spinal infections, post-operative complications, vascular malformation, and primary spinal tumors~\cite{balasubramanya2020lumbar}. 
As a result, registering spinal images from different modalities benefits various clinical contexts, ranging from improved diagnosis and proper treatment planning to personalized therapeutic decisions.

A major problem when registering articulated rigid structures is their relative motion and the soft tissue deformations due to patient movement. Indeed, rigid registration cannot address the problem of the varying curvature of patients' spine during different imaging sessions, as well as a global deformable registration ignores the difference between
soft tissues and bony structures. Consequently, these general-purpose methods are not fully appropriate to solve our problem.

Few works have addressed the problem of rigid structures in conventional deformable registration. \cite{little1997deformations} and \cite{bukas2021patient} defined a way to interpolate the deformations of rigid objects based on their \gls{EDT}. \cite{arsigny2005polyrigid} proposed a poly-rigid/affine transformation model, which produces locally rigid/affine diffeomorphic deformation field. \cite{loeckx2004nonrigid} and \cite{staring2007rigidity} imposed rigidity by introducing several constraints, including the linearity, orthonormality, and identity determinant of the transform matrix. \cite{gill2012biomechanically} defined groups of springs between vertebrae and penalized the change in length of them. In addition, \cite{kim2013distance} penalized the inter-voxel distance change in rigid bodies. Finally, \cite{rohlfing2003volume} proposed penalizing deviations of the Jacobian determinant of the deformation from unity to preserve the volume of lesions in breast MRI registration. \cite{staring2007rigidity} introduced the latter as one of the rigidity penalties in deformable registration, while proposing to use the orthonormality of a rigid transform as a rigidity penalty in registering digital subtraction angiography (DSA) images. However, these approaches are based on conventional iterative optimization methods, which are time-consuming, and their performance highly depends on the initialization and parameter settings.
In more recent years, \gls{CNNs} were used to do \gls{DLIR}~\cite{haskins2020deep}. Their main advantage is that they enable inferring a plausible transformation in a single iteration, using a model trained optimizing only image similarity and deformation smoothness~\cite{balakrishnan2019voxelmorph,de2019deep,mok2021conditional}.
%\cite{mok2020large} addressed large deformations by introducing Laplacian pyramid image registration networks (LapIRN). 
% Where biomechanical constraints are present, general-purpose methods fail.  
% For registration of MR-CT spine images, rigid registration cannot address the problem of the varying curvature of patients' spine during different imaging sessions. In contrast, a global deformable registration ignores the difference between soft tissues and bony structures.
A biomechanically-constrained method for \gls{DLIR} of MRI-CT images of the prostate, proposed by \cite{fu2021deformable}, consists of training on \gls{FE} modeling-generated motion fields. However, it required the segmentation of the prostate in both modalities to establish the surface point cloud correspondence. This segmentation is non-trivial for the spine in MRI.
% \jo{I would put this sentence above to explain why there is a problem at all.(solved)}
% For registration of MR-CT spine images, rigid registration cannot address the problem of the different curvature of patients' spine during different imaging sessions. In contrast, the global deformation field obtained by deformable registration ignores the difference between soft tissues and bony structures. 

Another hurdle for unsupervised \gls{DLIR} methods is the selection of the similarity metric for multi-modal image registration. For registration of MRI-CT of the pelvis, \cite{momin2021ct} introduced a new metric for unsupervised training of a deep network called self-correlation descriptor. \cite{mckenzie2020multimodality} reduced the problem of MRI-CT registration of images of the neck to a mono-modal one by training a neural network to synthesize CT images from MRI. \cite{hu2018weakly} proposed a weakly-supervised framework that utilizes the \gls{DSC} between the labels of the fixed image and the deformed moving image without any image-based similarity metrics during training. None of the above consider the rigidity of bony structures and the constraints that they impose on the deformation field. As a result, those registration results are not suitable for image fusion in the spine.

% uses the labels on training in addition to the images. By using the \gls{DSC} between the labels of the fixed image and the deformed moving image, no multi-modal similarity metric is needed. 
Our work introduces a weakly-supervised anatomy-aware method for registering spinal MR-CT images. We acknowledge that vertebrae segmentation is not a trivial task for \gls{MRI}, thus we only rely on label maps from \gls{CT} for training. Further, we devise losses and metrics to deal with the biomechanical constraints imposed by bony structures. Our main contributions are:%In this paper, we propose a weakly-supervised biomechanically-constrained registration framework for spine CT and T1-weighted MR images, which is aware of the rigidity of the vertebrae during registration. 

% We
% \begin{itemize}
%     \item introduce (one new loss for volume-) and two new losses for rigidity-preservation;
%     \item adapt two rigidity penalties in conventional registration to deep learning scenarios;
%     \item evaluate the influence of each loss on a in-house dataset with 167 patients
% \end{itemize}
% \paragraph{Contributions}
\begin{itemize}
    \item Proposal of a framework for rigidity-preserving MRI/CT deformable registration of the spine taking rigidly aligned CT/MRI images as input
    \item Introduction of the rigid dice loss and rigid field loss for rigidity-preservation
    \item Adaptation of rigidity penalties used in conventional registration of spine to \gls{DLIR} (orthonormal condition, properness condition)
    \item Extensive evaluation and ablation study of different losses on an in-house dataset with 167 patients
    % on two datasets, an in-house dataset with 150 patients and the dataset from Learn2Reg challenge
\end{itemize}

\section{Method}

\subsubsection{Architecture}
Let $\mathcal{F}: \Omega_f \rightarrow \mathbb{R}$ denote the fixed image, and $\mathcal{M}:\Omega_m \rightarrow \mathbb{R}$ denote the moving image, where $\Omega_f\in\mathbb{R}^3$ and $\Omega_m\in\mathbb{R}^3$ represent the coordinate space of $\mathcal{F}$ and $\mathcal{M}$. In our setup, the training loss consists of three terms. Firstly, the network learns to establish spatial correspondence between fixed and moving images by computing a \gls{DDF} $\phi:\Omega_f\rightarrow\Omega_m$ through an intensity-based image similarity loss $\mathcal{L}_{\text{sim}}$. Second, a smootheness regularizer $\mathcal{L}_{\text{smooth}}$ ensures the output \gls{DDF} is plausible and realistic. Lastly, we define rigidity penalties $\mathcal{L}_{\text{rigid}}$ between the moving label and the warped label, or on the deformation vectors inside the rigid bodies, to guarantee each of them is transformed rigidly. See their formal definition in the next subsubsections.

The network is trained by minimizing the following loss function:
\begin{equation}
    \mathcal{L}=\mathcal{L}_{\text{sim}}(\mathcal{F},\mathcal{M}\circ\phi)
    +\lambda_{\text{smooth}}\mathcal{L}_{\text{smooth}}(\mathbf{v})
    +\lambda_{\text{rigid}}\sum_i^N\mathcal{L}_{\text{rigid}}(s^i_\mathcal{M},s^i_\mathcal{M}\circ\phi,\phi)
\end{equation}
where $\mathcal{M}\circ\phi:\Omega_f\rightarrow\mathbb{R}$ represents the warped moving image, $s^i_{\mathcal{M}}:\Omega_m\rightarrow \{0,1\}$ is the binary segmentation label of the $i$th rigid body in the moving image, $N$ is the number of rigid bodies, i.e., in our case, the vertebrae, and $\mathbf{v}$ is the \gls{SVF} parameterizing the \gls{DDF} $\phi$. The smoothness regularizer $\mathcal{L}_\text{smooth}$ is computed as the $l2$-norm of the diffusion on the spatial gradient of the \gls{SVF} $\mathbf{v}$.

We adopt the diffeomorphic version of VoxelMorph~\cite{balakrishnan2019voxelmorph} as our baseline network. This 3D network architecture comprises a UNet~\cite{ronneberger2015u} and two convolutional layers with 32 filters each to output the \gls{SVF} $\mathbf{v}$. The UNet consists of an encoder with convolution filter channels $[32,32,64,64]$ and a decoder with channels $[64,64,64,64]$. To guarantee the invertibility and topology preservation, the predicted deformations are parameterized using the \gls{SVF} under the Log-Euclidean framework. $\mathbf{v}$ is integrated using the \textit{scaling and squaring} method~\cite{arsigny2006log} with $T=7$ time steps to obtain the diffeomorphic \gls{DDF} $\phi$~\cite{balakrishnan2019voxelmorph}.

\begin{figure}[!ht]
  \centering
  \includegraphics[width=0.85\textwidth]{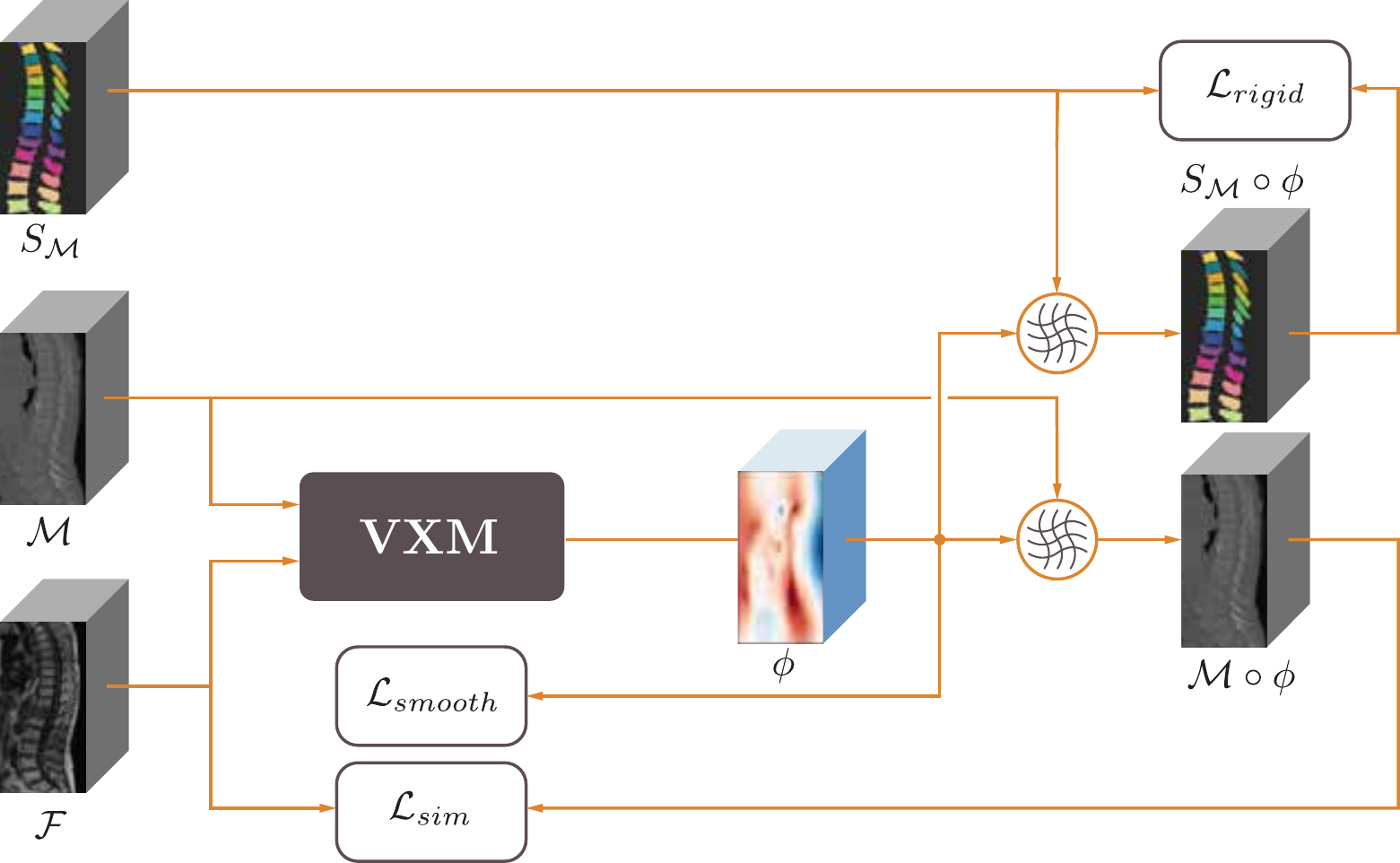}
  \caption{An overview of the architecture: A VoxelMorph network (VXM) takes as input a \gls{CT}  as moving image $\mathcal{M}$, its label map $\mathcal{S}_\mathcal{M}$, and an \gls{MRI} as fixed image $\mathcal{F}$. The output is a \gls{DDF} $\phi$. Three losses are applied on the warped \gls{CT} ($\mathcal{L}_{\text{sim}}$), the warped label map ($\mathcal{L}_{\text{rigid}}$), and the \gls{DDF} itself ($\mathcal{L}_{\text{smooth}}$).}
  \label{fig: network}
\end{figure}
% \subsection{Anatomy-aware losses}
% \subsubsection{direct volume loss}
% According to the property of rigid bodies, their volume will not change when they undergo transformation. We calculate the volume of each vertebra by summing its binary segmentation label, and compute the difference between moving label and warped label. By penalizing the change of volume during registration, the volume of each vertebra is preserved, thus contributing to the maintenance of the rigidity.
\subsubsection{Rigid dice loss}\label{sec:rigid_dice}
% The inferred transformation should preserve the structure of vertebrae. 
To penalize structural changes on rigid parts, i.e. the vertebrae, we use a rigid dice loss. Let $\phi$ be the inferred deformation, then for each vertebra in the moving image ($v_i$) with binary segmentation label $s^i_{\mathcal{M}}$, we calculate the closest rigid transform ($T_{\text{rigid}}^{i,\phi}$) by solving the equation:
\begin{equation}
    T_{\text{rigid}}^{i,\phi} = \argmin_{T_{\text{rigid}}} \left( s^i_{\mathcal{M}}\circ T_{\text{rigid}}- s^i_{\mathcal{M}}\circ \phi \right)^2.
\end{equation}
Then the rigid dice loss for each vertebra
% ($\mathcal{L}_{rd}$) 
can be defined as:
\begin{align}
    \mathcal{L}^i_{\text{rigid dice}} &= 1 - 2*\frac{\left|(s^i_{\mathcal{M}}\circ T_{\text{rigid}}^{i,\phi}) \cap (s^i_{\mathcal{M}}\circ \phi)\right|}
    {\left|s^i_{\mathcal{M}}\circ T_{\text{rigid}}^{i,\phi}\right|+\left|s^i_{\mathcal{M}}\circ \phi\right|}
    % \nonumber \\
    % \mathcal{L}_{rd} &= \frac{1}{N} \sum_i^N \mathcal{L}^i_{\text{Dice}}
    % {\left|s^i_{\mathcal{M}}\circ T_{\text{rigid}}^{i,\phi}\right|+\left|s^i_{\mathcal{M}}\circ \phi\right|}
    % \frac{1}{N} \sum_i^N \text{Dice}\left( s^i_{\mathcal{M}}\circ T_{\text{rigid}}^{i,\phi} ,s^i_{\mathcal{M}}\circ \phi \right).
\end{align}

\subsubsection{Rigid field loss}
To enforce a rigid transformation for each rigid body, we evaluate if the deformation field within the rigid body is close to a rigid deformation using the rigid field loss. Let $P_i=\{\mathbf{p}_j\}_{j=1,...,n},\mathbf{p}_j\in s^i_{\mathcal{M}}$ be a set of randomly selected points from $i$th vertebra ($v_i$) in the moving image and $Q_i=\{\mathbf{q}_j\}_{j=1,...,n},\mathbf{q}_j\in s^i_{\mathcal{M}}\circ\phi$ the set of corresponding points to $P_i$ in the warped image. Using $P_i$ and $Q_i$, we can compute the average rigid transform for $v_i$ ($\phi^i_{\text{rigid}}$) by solving a least squares problem with singular value decomposition (SVD)~\cite{sorkine2017least}.
%as follows:
% 
% \begin{align}
%     U_i, \Sigma_i, V_i^T &= \text{SVD}\left((Q_i - \Bar{Q_i})^T(P_i - \Bar{P_i})\right) \nonumber \\
%     R_i &= V_i U_i^T \nonumber \\
%     \mathbf{t}_i &= \Bar{Q_i}-R_i\Bar{P_i} \\
%     \phi^i_{\text{rigid}} &= R_i\mathbf{z}+\mathbf{t}_i,  \forall\mathbf{z}\in\Omega_{\mathcal{F}} \nonumber
% \end{align}
% 
% where ... Then we can calculate loss for each vertebra and the overall loss:
% 
% \begin{align}
%     \mathcal{L}^i_{\text{field}} &= \sum (s^i_{\mathcal{M}}\circ\phi)(\phi^i_{\text{rigid}}-\phi)^2 /
% \left|s^i_{\mathcal{M}}\circ\phi\right| \nonumber \\
%     \mathcal{L}_{rigid\ field} &= \frac{1}{N}\sum_i^N\mathcal{L}^i_{\text{field}} 
% \end{align}
% 
% Since the \gls{DDF} defines the spatial correspondence between the moving image $\mathcal{M}$ and the warped image $\mathcal{M}\circ\phi$, when a set of points is sampled inside a vertebra in the moving image, the corresponding points in the warped image can be easily obtained through the \gls{DDF}. With the set of correspondences in each vertebra, we can compute the average rigid transform by solving least squares problem with singular value decomposition (SVD)~\cite{sorkine2017least}. 
Then, by minimizing the distance between the predicted deformation vectors and the average rigid transform vectors inside each vertebra, the network learns to move it rigidly (see also Supplementary Material).
\begin{equation}
    \mathcal{L}^i_{\text{rigid field}} = 
\sum_{\mathbf{z}\in s^i_{\mathcal{M}}\circ\phi}(\phi^i_{\text{rigid}}(\mathbf{z})-\phi(\mathbf{z}))^2 /
\left|s^i_{\mathcal{M}}\circ\phi\right|
\end{equation}

\subsubsection{Properness condition}
% \cite{rohlfing2003volume} proposed penalizing deviations of the Jacobian determinant of the deformation from unity to preserve the volume of lesions in breast MRI registration. \cite{staring2007rigidity} introduced it as one of the rigidity penalties in deformable registration and called it \gls{PC}.
We adapt the penalty of the determinant of the Jacobian proposed by \cite{rohlfing2003volume} to the \gls{DLIR} scenario and call it \gls{PC} as proposed by \cite{staring2007rigidity}. To implement it, first, we compute the ideal rotation matrix $R$ of every voxel $\mathbf{z}\in\Omega_{\mathcal{F}}$ from the Jacobian of the \gls{DDF} $\phi$. 
\begin{equation}
    R(x) = J_\phi(x)
\end{equation}
Then, we constrain the rotation matrices $R(x)$ of voxels inside each vertebra to have a proper unity determinant by minimizing the $l2$-distance between the Jacobian determinant and constant one:
\begin{equation}
    \mathcal{L}_{\text{pc}}=\frac{1}{N}\sum_i^N\frac{1}{\left|s^i_{\mathcal{M}}\circ\phi\right|}\sum_{x\in s^i_{\mathcal{M}}\circ\phi}\left\|\det J_\phi(x)-1\right\|_2^2
\end{equation}

\subsubsection{Orthonormal condition}
% \cite{staring2007rigidity} also proposed using the orthonormality of a rigid transform as a rigidity penalty in registering digital subtraction angiography (DSA) images, i.e. the \gls{OC}.  
We include the \gls{OC} proposed by \cite{staring2007rigidity} in our setup, by computing the inner product of the Jacobian of the \gls{DDF} $J_\phi$ and by penalizing the deviation of it from an identity matrix using the matrix norm. By forcing the rotation $R(x)$ of every voxel inside each vertebra to be orthonormal, the rigidity is preserved:
\begin{equation}
    \mathcal{L}_{\text{oc}}=\frac{1}{N}\sum_i^N\frac{1}{\left|s^i_{\mathcal{M}}\circ\phi\right|}\sum_{x\in s^i_{\mathcal{M}}\circ\phi}\left\|J_\phi(x) ^T J_\phi(x) - I\right\|_{\text{fro}}^2
\end{equation}

% \subsection{Initialization}

% Extraction of center points

\section{Experiments}

\subsubsection{Dataset}
We use an in-house dataset of $167$ patients. The spatial resolution of the CTs ranges from $(0.2,0.2,0.4)$ mm to $(1.5,1.5,5)$ mm, while the T1-weighted MRIs have voxel spacing from $(0.3,0.3,2.7)$ mm to $(1,1,5)$ mm. The dataset has different fields of views and covers different part of spines, which in total, results in images of $1280$ vertebrae. We split our dataset into training, validation and test sets with $117$, $25$ and $25$ patients in each set, respectively. During training and inference, all images are resampled to $1$ mm isotropic resolution, and the intensities are normalized to $[0,1]$. Vertebra detection and segmentation on CT images is done automatically using the framework of~\cite{sekuboyina2018btrfly}.
For MRI, an expert annotated the central point of each vertebra, and manually segmented the validation and test set. The vertebral central points of both modalities are used to rigidly register each image pair. The ground truth segmentation labels are used for measuring the registration accuracy in terms of \gls{DSC}.

\subsubsection{Hyperparameter tuning of image-similarity and smoothness}
We investigated three different multi-modal image similarity losses: \gls{NMI}~\cite{wells1996multi}, \gls{NGF}~\cite{pluim2000image,haber2006intensity}, and a \gls{MIND}-based~\cite{heinrich2013towards} loss. The \gls{MIND}-based loss achieved the highest validation \gls{DSC}. The hyperparameter $\lambda_\text{smooth}$ was optimized with respect to the \gls{MIND}-based similarity loss based on the validation set. Considering both the registration accuracy (given by the \gls{DSC}) and the validity of the inferred transformation (given by the the standard deviation (SD) of the logarithm of the Jacobian determinant, $\text{SD}\log|J_\phi|$), we find $0.01$ to have the best performance so $\lambda_\text{smooth}$ is fixed to $0.01$ for the rest of the experiments. See validation results of the tuning in the Supplementary Material.
% In our proposed method we have two hyperparameters, $\lambda_{\text{smooth}}$ and $\lambda_{\text{rigid}}$. We first optimize for $\lambda_{\text{smooth}}$ without including the rigidity loss. Based in our experiments, the acceptable range for $\lambda_{\text{smooth}}$ is between $[0.01 \ 0.05]$ and we set it to $0.01$ for the rest of the experiments.

% Since we have multiple options for the rigidity loss, we optimize $\lambda_{rigid}$ for each one separately. The results of these experiments can be seen in \checktext{table}.

\subsubsection{Ablation Studies}
To measure the effect of each loss we employed for rigidity, we perform an ablation study for each of them. First, we train our network without any rigidity penalties ($\lambda_\text{rigid}=0$) for 500 epochs to get our baseline model. Then, starting from the milestone at 400 epochs, we add each of the rigidity penalties and train for another 100 epochs. We also compare our method with one conventional approach~\cite{staring2007rigidity} (denoted as \textit{Staring}), which takes \gls{NMI} with \gls{OC} and \gls{PC} as loss function and iteratively optimizes for each pair of data.
\begin{figure}[ht]
  \centering
  \includegraphics[width=\textwidth]{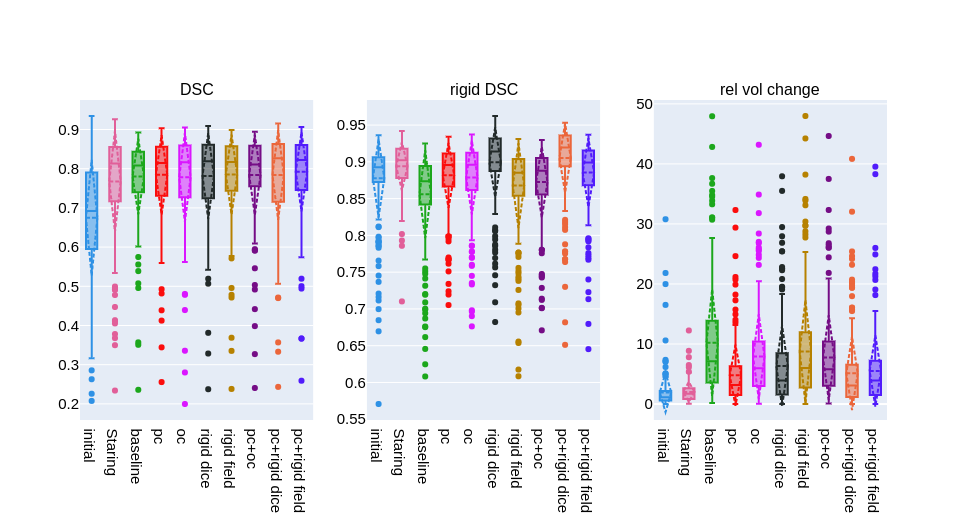}
  \caption{Boxplots depicting the \gls{DSC}, rigid dice and relative volume change $\%\Delta$vol of initial dataset, conventional method, baseline and each loss setting.}
  \label{fig: boxplot}
\end{figure}
\begin{figure}[ht!]
    \centering
    \subfloat[$\mathcal{M}$]{\includegraphics[height=0.16\textheight]{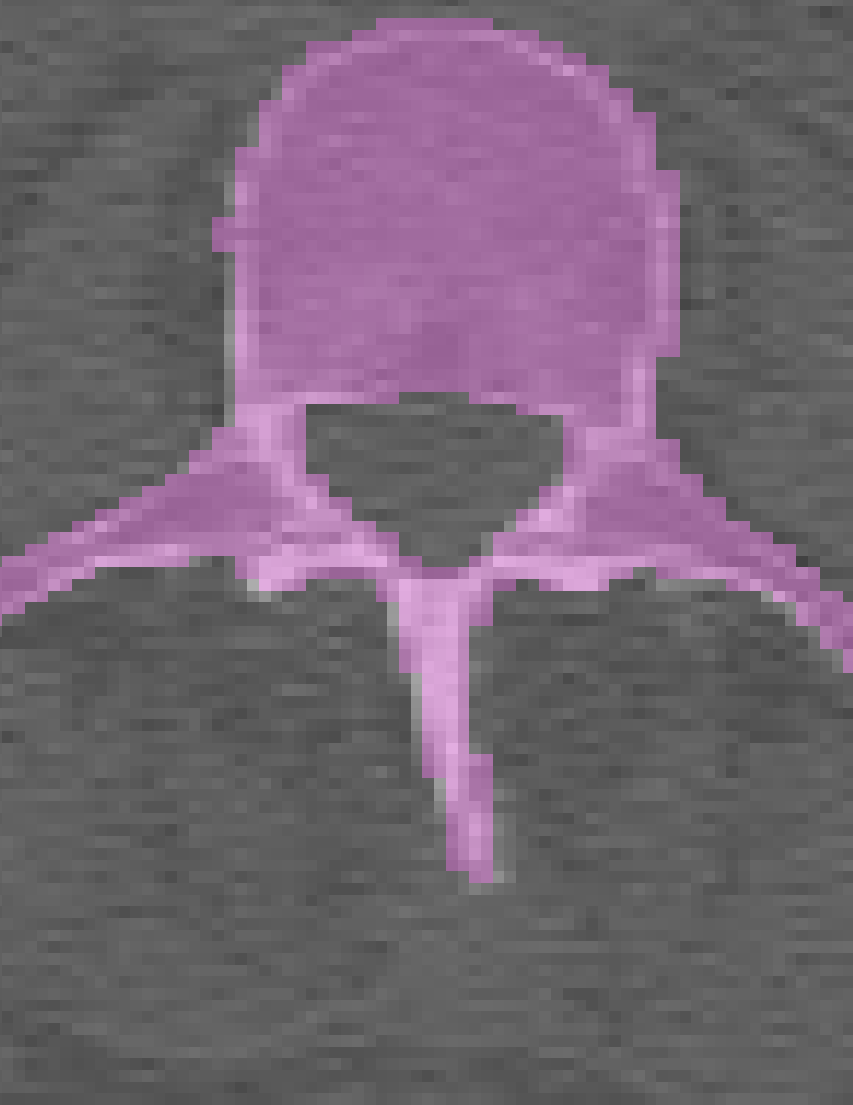}}
    \subfloat[con]{\includegraphics[height=0.16\textheight]{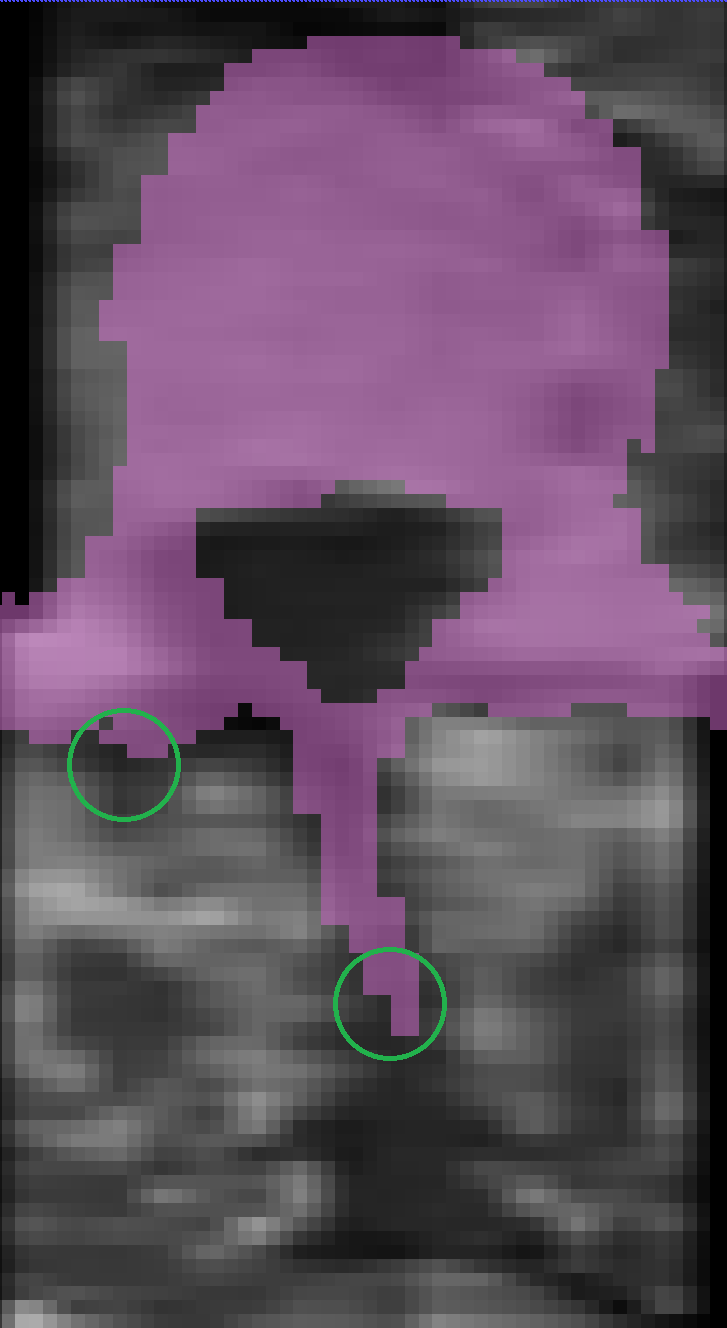}}
    \subfloat[baseline]{\includegraphics[height=0.16\textheight]{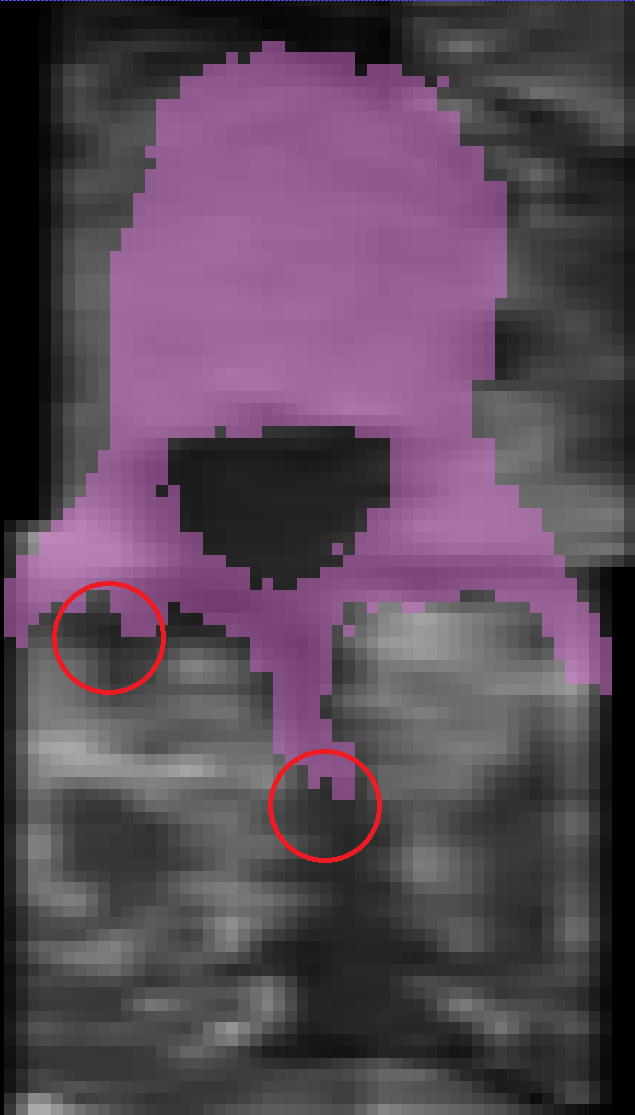}}
    \subfloat[rig dice]{\includegraphics[height=0.16\textheight]{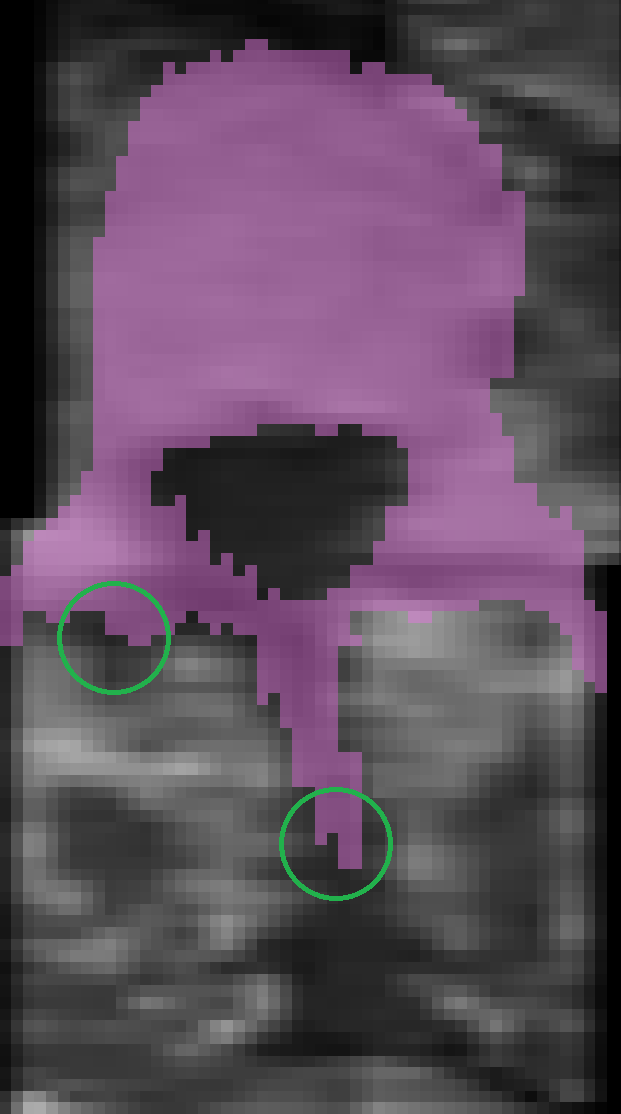}}
    \subfloat[pc+oc]{\includegraphics[height=0.16\textheight]{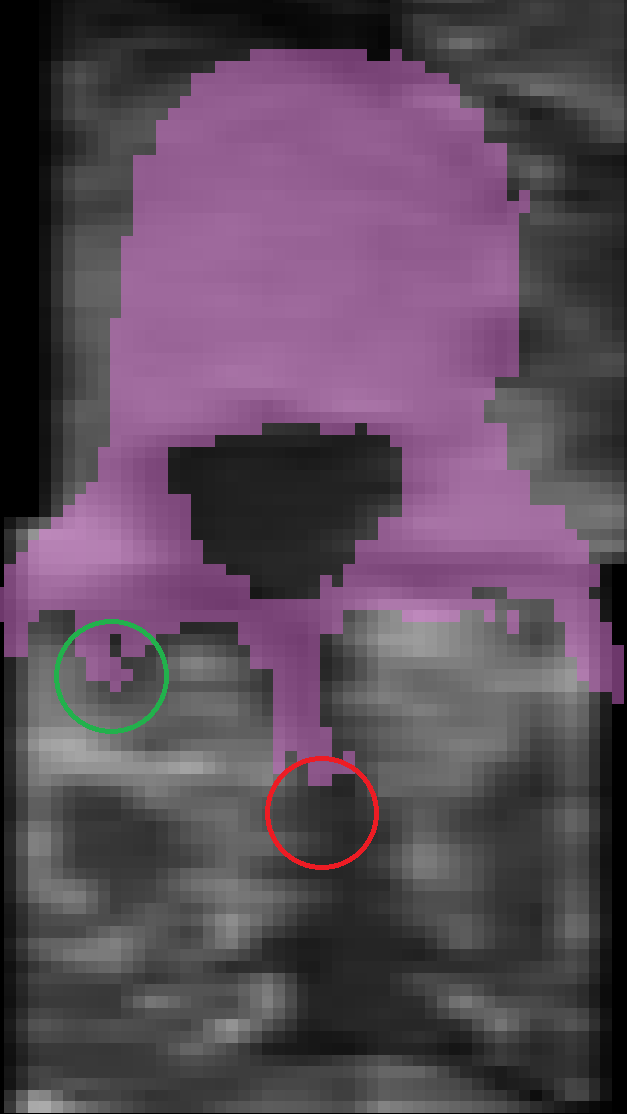}}
    \subfloat[$\mathcal{F}$]{\includegraphics[height=0.16\textheight]{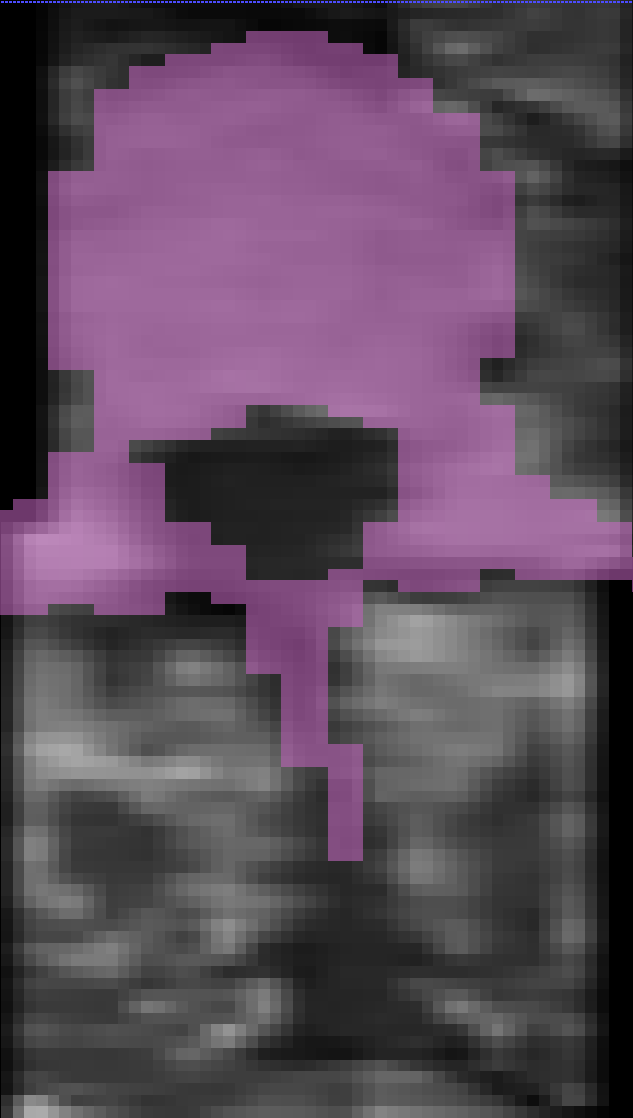}}\\
    \subfloat[$\mathcal{M}$]{\includegraphics[height=0.15\textheight]{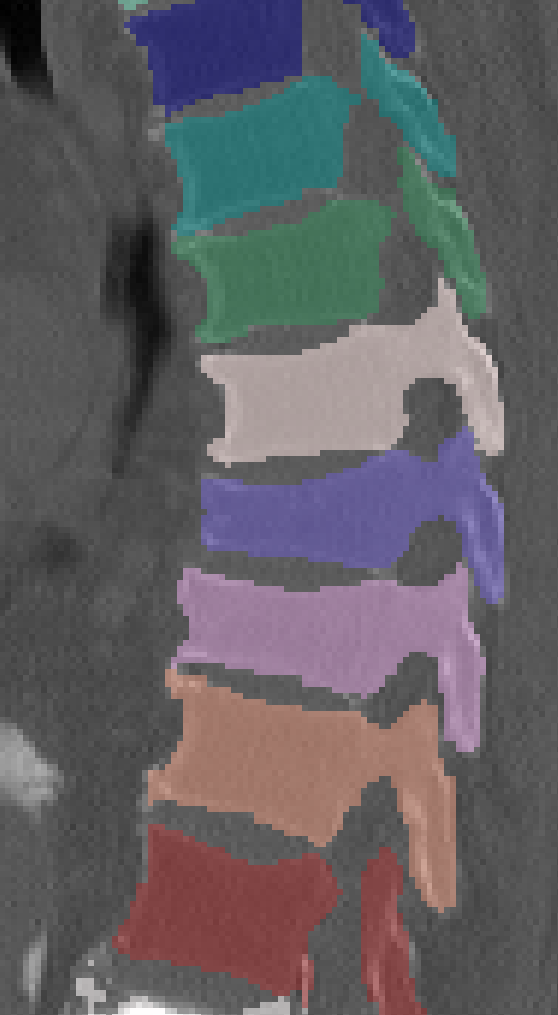}}
    \subfloat[con]{\includegraphics[height=0.15\textheight]{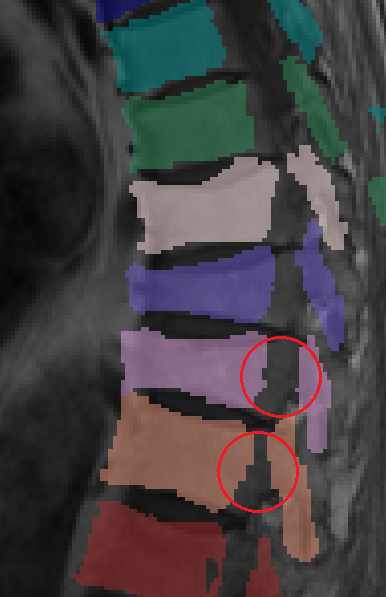}}
    \subfloat[baseline]{\includegraphics[height=0.15\textheight]{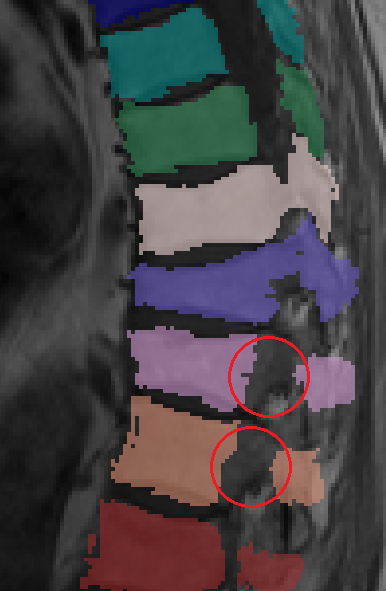}}
    \subfloat[rig dice]{\includegraphics[height=0.15\textheight]{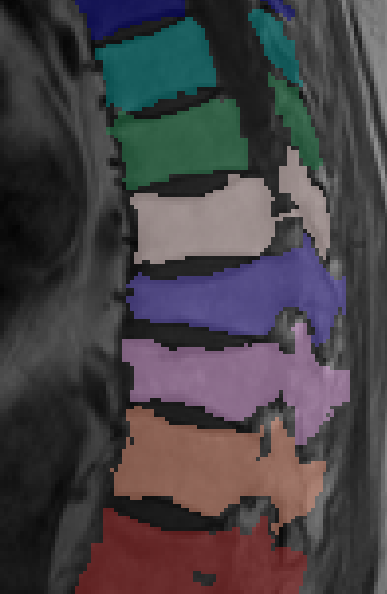}}
    \subfloat[pc+oc]{\includegraphics[height=0.15\textheight]{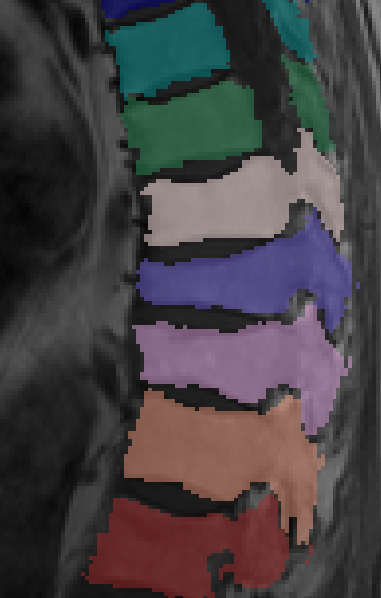}}
    \subfloat[$\mathcal{F}$]{\includegraphics[height=0.15\textheight]{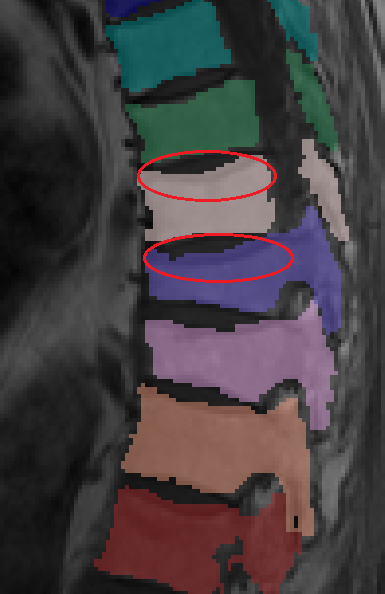}}
    \caption{Example axial (top) and sagittal (bottom) slices of moving image CT, fixed image superimposed by warped label from conventional, baseline, and other methods with different rigidity penalties.}
    \label{fig:vis_example}
\end{figure}
\subsubsection{Results}
Table~\ref{tab:test_metrics} gives a detailed summary of our experiments (see Supplementary Material for weights setting). The baseline method performs better than the conventional \textit{Staring} method, with $1.5\%$ improvement in \gls{DSC} and $100\times$ faster speed. However, it cannot guarantee the rigidity of the vertebrae and their volume drastically changes during registration. Our proposed anatomy-aware losses alleviate this issue. Specifically, $\mathcal{L}_\text{pc}$ contributes significantly to the volume preservation, with only $4.8\%$ loss per vertebra (Fig.~\ref{fig:vis_example}). The $\mathcal{L}_{\text{rigid dice}}$, $\mathcal{L}_{\text{rigid field}}$ and $\mathcal{L}_{\text{oc}}$  attain higher rigid dice than the baseline method, among which the rigid dice loss achieves the highest level of rigidity, with $5.7\%$ improvement over the baseline. The combinations of \gls{PC} and other penalties reduce the volume loss while maintaining the rigidity of the vertebrae, and yielding plausible and smooth \gls{DDF}. Fig.~\ref{fig: boxplot} shows the boxplots of our experiments. The outliers are the cases where the boundary vertebrae are incomplete in CT/MRI images. With comparable registration accuracy, the proposed anatomy-aware losses significantly preserve the volume and rigid properties of the vertebrae. Qualitative results of the different methods are displayed in Fig.~\ref{fig:vis_example}, where the red circles are showing regions of incorrect warped labels, while the greens outline correct ones.
In particular, the axial views show that training with rigid dice loss help maintain the details in the process area. The sagittal views validate that the \gls{PC} preserves the volume, and the \gls{OC} results in sharper borders. As shown in the axial slices, the incomplete transverse processes in MR images limit our method from getting high rigid \gls{DSC}, since the network cannot predict missing parts.

\subsubsection{\gls{PC} during inference} Finally, to validate \gls{PC}'s effectiveness in volume preservation, we compare its performance with the direct volume loss, which tries to penalize the volume change before and after the registration directly. We also compute \gls{PC} as a metric (lower is better), during inference (see Supplementary Material). With a similar $\%\Delta\text{vol}$, the direct volume loss gets a higher \gls{PC} than training with \gls{PC} in the test set. It indicates that \gls{PC} preserves the volume in a more realistic way, without any compression or expansion in compensation. Moreover, training with \gls{OC} can also decrease \gls{PC}, which implies that enforcing the orthonormality of the transformation also contributes to unity determinant.

\begin{table}[h]
    \centering
    \caption{Quantitative summary of the test set results given as averages(SD). \gls{DSC} indicates the mean Dice score over all vertebrae. Similar to $\mathcal{L}_{\text{rigid dice}}$ (Sec~\ref{sec:rigid_dice}), rigid DSC computes the \gls{DSC} between the warped label $s^i_\mathcal{M}\circ\phi$ and the rigidly transformed label $s^i_\mathcal{M}\circ T_\text{rigid}^i$, to measure the level of rigidity kept during registration. $\%\Delta$vol represents the relative volume change of each vertebra between source label $s_\mathcal{M}^i$ and warped label $s_\mathcal{M}^i\circ\phi$. To assess the plausibility of the inferred transformation, $|J_\phi|_{\leq0}$ indicates the number of folding voxels in the \gls{DDF} (less is better), $\text{SD}\log|J_\phi|$ measures the smoothness of the \gls{DDF} (lower is better). The runtime is measured with CPU and with second as unit.\\}
    \begin{tabular}{ccccccc}
    \Xhline{1pt}
%    \multirow{2}{*}{Method} & \multicolumn{6}{c}{Test set} \\ \cline{2-7}
     Method                       & DSC & Rigid DSC & $\%\Delta\text{vol}$ & $|J_\phi|_{\leq0}$ & $\text{SD}\log|J_\phi|$ & Time\\
    \Xhline{0.8pt}
    Initial  & 0.67(0.15) & 0.88(0.05) & 4.16(3.95) & - & - & -  \\
    \Xhline{0.8pt}
    Staring & 0.77(0.13) & 0.89(0.03) & 2.02(1.65) & \textbf{0} & \textbf{0.04(0.01)} & 243 \\
    Our baseline & 0.78(0.10) & 0.86(0.06) & 10.21(8.91) & 0.04(0.20) & 0.13(0.05) & 3.5(1.3) \\
    \Xhline{0.8pt}
    pc     & 0.78(0.10) & 0.88(0.04) & \textbf{4.80(5.08)} & 1.60(7.84) & 0.12(0.05) & 3.7(1.2) \\
    oc     & 0.78(0.11) & 0.88(0.05) & 7.90(7.09) & 0.12(0.59) & 0.13(0.05) & 3.4(1.1) \\
    rigid dice  & 0.78(0.12) & 0.90(0.5) & 6.32(6.77) & 0.64(3.14) & 0.13(0.05) & \textbf{3.3(1.1)} \\
    rigid field  & 0.79(0.10) & 0.87(0.06) & 8.74(8.59) & 0.08(0.39) & 0.13(0.05) & 3.6(1.2) \\
    pc+oc     & 0.78(0.10) & 0.87(0.05) & 7.73(7.08) & \textbf{0} & 0.12(0.05) & 4.0(1.3) \\
    pc+rigid dice & 0.78(0.11) & \textbf{0.90(0.05)} & 5.12(6.13) & 18.30(63.20) & 0.13(0.06) & 3.4(1.1) \\
    pc+rigid field & \textbf{0.79(0.10)} & 0.89(0.05) & 5.51(5.94) & 2.40(11.80) & 0.13(0.06) & 3.6(1.2) \\
    \Xhline{1pt}
    \end{tabular}
    \label{tab:test_metrics}
\end{table}

\section{Conclusion}
In this paper, we present a framework for rigidity- and volume- preserving deformable registration of spinal CT/MRI. Compared to supervised \gls{DLIR} methods which need both labels of the fixed and moving images, our approach only requires the moving label. We define two novel rigidity penalties (rigid dice loss and rigid field loss) to constrain the movement of the vertebrae. Moreover, we introduce \gls{PC} and \gls{OC} to the \gls{DLIR} scenario. The results of extensive experiments indicate that \gls{PC} is competent in volume preservation while the other three penalties guarantee the rigid properties of the vertebrae during deformable registration. Compared to conventional methods, our algorithm achieves higher accuracy and is significantly faster. Moreover, the proposed penalties can be easily transferred to other \gls{DLIR} methods. %\checktext{With labels of rigid bodies provided, other pretrained network can also learn to preserve volume and rigidity through training with our losses for several epochs.} 
% Moreover, the proposed framework is highly generalizable to the deformable registration of other articulated rigid structures.
% \subsubsection{Acknowledgements} Please place your acknowledgments at
% the end of the paper, preceded by an unnumbered run-in heading (i.e.
% 3rd-level heading).

%\newpage
\bibliographystyle{splncs04}
\bibliography{ms.bib}
\end{document}